\documentclass[aps,showpacs,twocolumn]{revtex4}
\usepackage{amsfonts}
\usepackage{amsmath}
\usepackage{amssymb}
\usepackage{hyperref}
\usepackage{graphicx}
\usepackage{dcolumn}
\usepackage{amsmath}
\usepackage{grffile}
\usepackage{chngcntr}

\renewcommand{\k}{\mathbf{k}}

\begin{document}

\title{Electron Spectrum Topology and Giant  Density-of-States Singularities in Cubic Lattices}

\author{P.A.~Igoshev and V.Yu.~Irkhin}
\affiliation
{620108 M.N. Mikheev Institute of Metal Physics,  Ekaterinburg, Russia,\\
620002 Ural Federal University, Yekaterinburg, Russia}


\begin{abstract}
The topology of isoenergetic surfaces in reciprocal space for simple~(sc), body-centered~(bcc), and face-centered~(fcc) cubic lattices is investigated in detail in the tight-binding approximation, taking into account the transfer integrals between the nearest and next neighbors $t$ and $t'$. It is shown that, for values $\tau = t'/t = \tau_\ast$ corresponding to a change in the topology of surfaces, lines and surfaces of $\k$-van Hove points can be formed. With a small deviation of $\tau$ from these singular values, the spectrum in the vicinity of the van Hove line (surface) is replaced by a weak dependence on~$\k$ in the vicinity of several van Hove points that have a giant mass proportional to $|\tau - \tau_ \ast|^{-1}$. Singular contributions to the density of states near peculiar $\tau$ values are considered, analytical expressions for the density of states being obtained in terms of elliptic integrals. It is shown that in a number of cases the maximum value of the density of states is achieved at energies corresponding not to $\mathbf{k}$-points on the Brillouin zone edges, but to its \textit{internal} points in highly symmetrical directions. The corresponding contributions to electron and magnetic properties are treated, in particular, in application to weak itinerant magnets.
\end{abstract}

\maketitle

\section{Introduction}
Van Hove singularities in the electron spectrum, in particular, in the electron density of states (DOS) $\rho(\epsilon)$ as a function of the energy $\epsilon$, are responsible for some specific features of electronic and magnetic properties. The corresponding singular contributions to the thermal characteristics promote structural and magnetic phase transitions~\cite{1993:Trefilov}. The geometric origin of electron peaks in the DOS was studied in detail in~Refs.~\cite{1993:Trefilov,1990:Peschanskih}. Simple examples of such systems are the bcc phase of Ca and the fcc phase of Sr, where the one-dimensional manifolds with a weak dispersion (van Hove lines) at the faces of the Brillouin zone are located near the Fermi level. In bcc Ca, they approximately correspond to parts of the P--N and N--H lines, whereas in fcc Sr, they correspond to the X--U, U--L, L--K, K--U, and K--W lines. The analysis~\cite{1993:Trefilov} shows that the $D$ line in the bcc lattice of Li, V, Cr, Fe, and Ba usually looks like a van Hove line~\cite{1978:Moruzzy,1986:Papacostantopoulos}.

The L point at the face of the Brillouin zone, which corresponds to a large effective mass (for the spectrum calculated in the paramagnetic phase), determines the pronounced magnetism of nickel (the ``van Hove magnet'') and the behavior of its susceptibility above the Curie point~\cite{2017:Katanin}. 
\begin{figure}[t!]
\noindent
\includegraphics[width=0.45\textwidth]{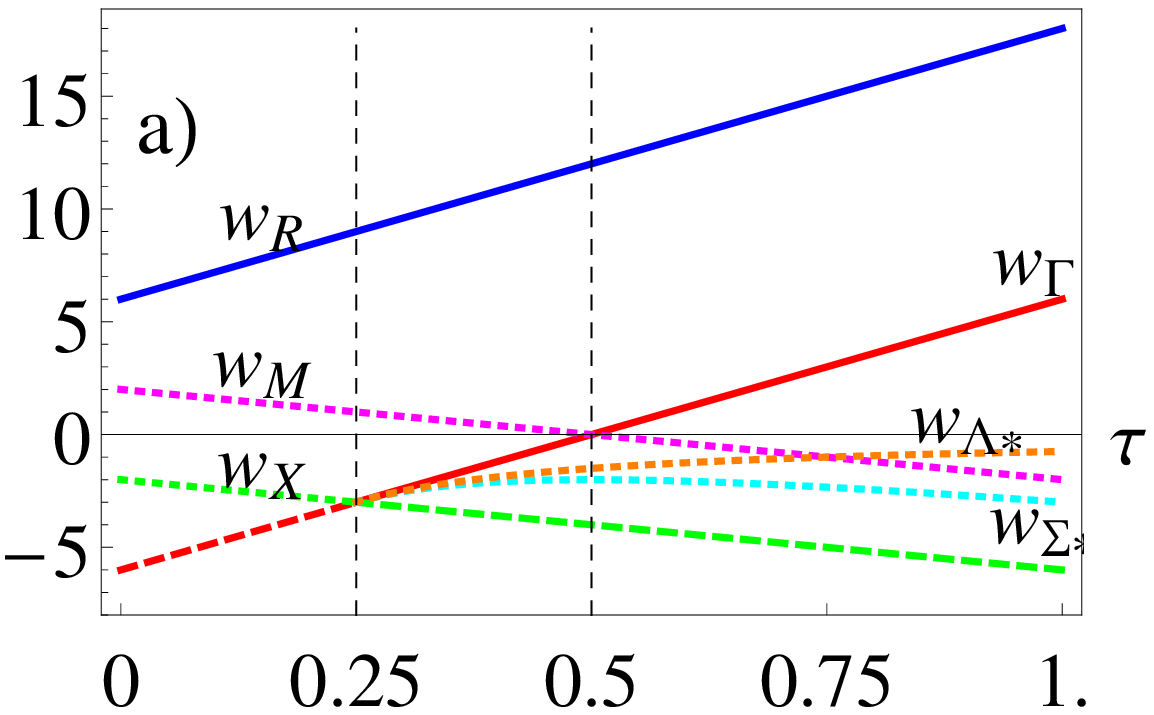}
\includegraphics[width=0.45\textwidth]{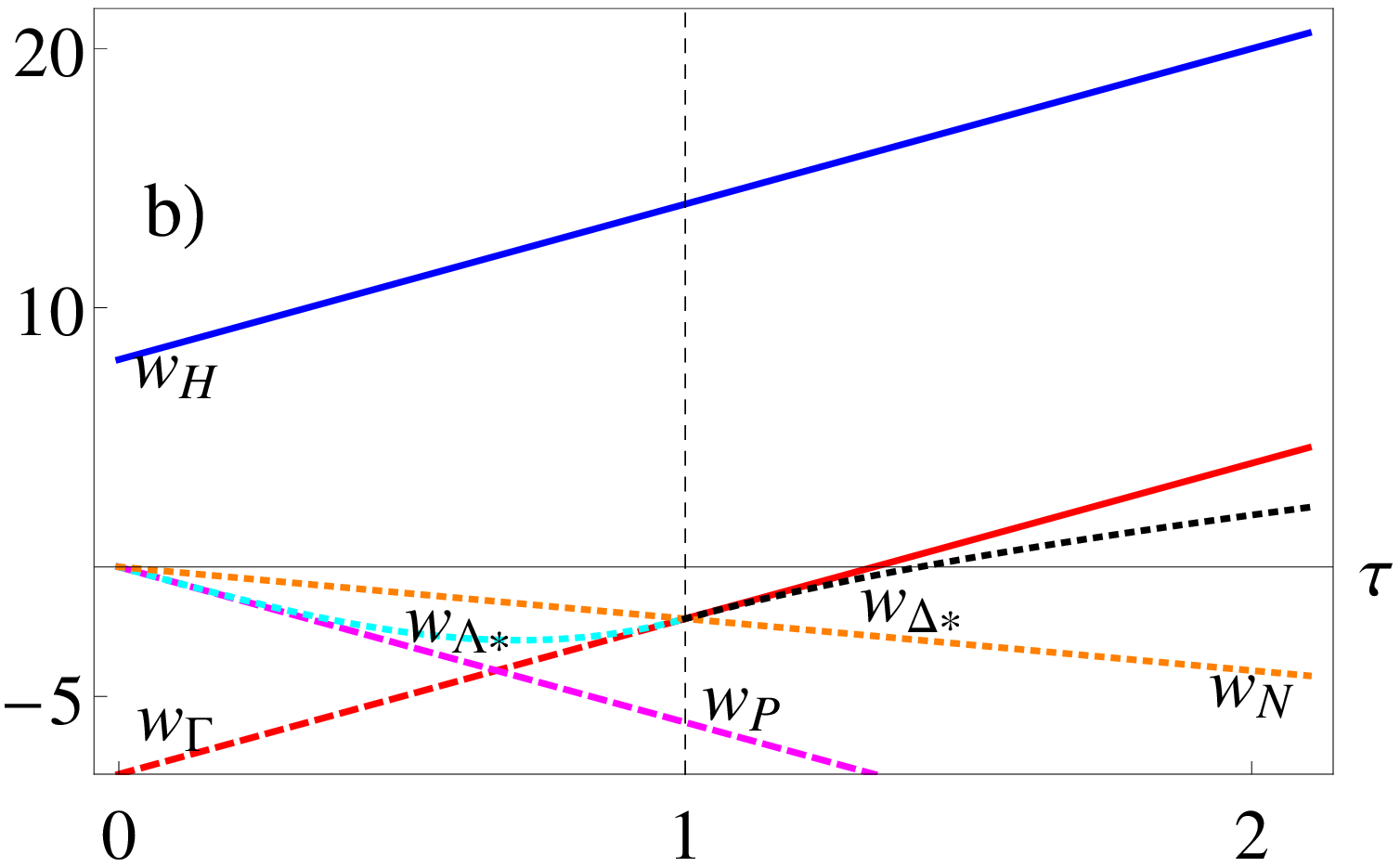}
\caption{$\tau$-dependence of the van Hove levels $w$ for the (a) sc and (b) bcc lattices. The solid, long-dashed, and short-dashed lines correspond to the local maximum, local minimum, and saddle point, respectively.
\label{fig:w}
} 
\end{figure}

The pronounced features in the DOS as a function of energy arise near the closely located van Hove $\k$ points (van Hove structures). The effect of such structures on the electron and lattice characteristics, including many-electron screening anomalies, was considered in~Refs.~\cite{Katsnelson_Trefilov21,Katsnelson_Trefilov22,1993:Trefilov}. The formation of narrow van Hove DOS peaks leads to the localization of electron states and enhances correlation effects.

This work is aimed at the detailed topological analysis of energy surfaces for cubic lattices with allowance for van Hove lines and surfaces and at the further application of such analysis to revealing the manifestation of topological features in the physical characteristics.

\section{Analysis of van Hove singularities}
A general relationship exists between the singularities in the density of states and the topological characteristic of the excitation spectrum $t(\k)$~\cite{1953:vanHove}. A point $\k$ in the reciprocal space is called a van Hove point if the velocity of elementary excitations satisfies the condition $\mathbf{v}(\mathbf{k}) = \partial t(\mathbf{k})/\partial\mathbf{k} = 0$. Singularities in the $\rho(\epsilon)$ plot are due to only energy levels $\epsilon = t(\mathbf{k})$ corresponding to van Hove points. The type of this singularity corresponding to an isolated van Hove point $\k$ is given by the signature of the quadratic form corresponding to the matrix $\partial^2 t(\mathbf{k})/\partial k_i\partial k_j$ (the signature is denoted as a number in slashes) that determines the behavior of the spectrum near this point (see details and definitions in~the~Appendix~\ref{sec:appendix}). Thus, a particularly important contribution to the DOS comes from van Hove points with a giant mass or from the merging of van Hove k points, which is closely related to an increase in one or more masses, as will be seen from the further analysis.

If the spectrum obtained by \textit{ab~initio} calculations exhibits only a slight dispersion within a certain range along some direction in the $\k$ space, then either a~point corresponding to a large mass or a pair of such points exists near this range.

We assume that hopping occurs only between the sites in the first (integral~$t$) and second ($t' = \tau t$) coordination spheres: $t(\mathbf{k}; \tau)$. 
Since $\rho(-\epsilon; -\tau) = \rho(\epsilon; \tau)$ for the bipartite lattices, below we assume that   $t = 1$, $\tau \ge 0$.

The explicit expressions for the DOS and lattice Green's functions were obtained earlier for a number of three-dimensional lattices taking into account only the nearest neighbors. This gives a symmetric function of energy $\rho(\epsilon, 0)$ for bipartite (sc and bcc) lattices~\cite{1969:Jelitto,1971:Katsura}. 
However, in a number of problems in modern condensed matter physics, the effects of asymmetry and next-nearest-neighbor hoppings are crucial (for example, in the electron spectrum of superconducting cuprates and for the metal-insulator transition in the antiferromagnetic phase~\cite{2019:Igoshev}); the corresponding results for the DOS are illustrated in the Appendix~\ref{sec:appendix}. 

For the sc lattice, the topology of the electron spectrum changes significantly with an increase in $\tau$ above $\tau_\ast^{\rm sc} = 1/4$~(at this value, the $\k$ points form a whole van Hove line~$\varDelta$; the $\Gamma$ point changes its type from minimum to maximum, and the X point changes from a saddle point to a local minimum) (see~Fig.~\ref{fig:w}a and Table~\ref{table:sc} in~the~Appendix). 
At $\tau < \tau_\ast^{\rm sc}$, there exist four van Hove points corresponding to the minimum at $\Gamma$, maximum at R, and two saddle points with opposite signatures, X/-1/ and M/-1/. 
Between them, the $\rho_{\rm sc}(\epsilon, \tau)$ plot as a function of $\epsilon$ has a wide plateau. 
At $\tau > \tau_\ast^{\rm sc}$, two additional van Hove $\k$ points, which lie in the $\varSigma$ and $\varLambda$ lines, are separated from~$\Gamma$ point (see Table~\ref{table:sc} in~the~Appendix); we denote these points as $\varSigma^\ast$ and $\varLambda^\ast$. 
With a further increase in $\tau$, these points move away from~$\Gamma$~point. 
For each of these points, all three masses diverge as $|\tau-\tau_\ast^{\rm sc}|^{-1}$ at $\tau\rightarrow\tau_\ast^{\rm sc}$. Thus, at $\tau = \tau_\ast^{\rm sc}$, the van Hove line~$\varDelta$ is split into three adjacent van Hove points $\Gamma$, $\varSigma^\ast/+1/$, $\varLambda^\ast/-1/$. 
At $\tau > \tau_\ast^{\rm sc}$, the $\rho(\epsilon, \tau)$ curve exhibits a quasisymmetric plateau between energies $w^{\rm sc}_{\varSigma^\ast}, w^{\rm sc}_{\varLambda^\ast}$; this plateau has the height of the order of $(\tau - \tau_\ast^{\rm sc})^{-1/2}$, the width of the top of $4(\tau - \tau_\ast^{\rm sc})^2/\tau$, and the universal dependence on the relative position of the energy level at this plateau (see Figs.~\ref{fig:DOS_vs_e}a and \ref{fig:DOS_vs_tau}a in the Appendix).

The change in the DOS at the top of the plateau is rather small because all three masses diverge at the same rate for both $\mathbf{k}$ points $\varSigma^\ast$ and $\varLambda^\ast$. To the right of the plateau, there is a sharp drop in $\rho(\epsilon, \tau)$ between the $w^{\rm sc}_{\varLambda^\ast}$ and $w^{\rm sc}_{\Gamma}$ levels. However, the difference between these levels is much larger than the width of the plateau. $\tau > \tau_\ast^{\rm sc}$, the maximum of the DOS is achieved at $\epsilon = w^{\rm sc}_{\varSigma^\ast}$ or $w^{\rm sc}_{\varLambda^\ast}$ (see Fig.~\ref{fig:DOS_vs_tau}a in the Appendix) corresponding to the internal van Hove $\k$ points of the Brillouin zone.

At $\tau = \tau_\ast^{\rm sc}$, the plateau completely disappears, and the giant van Hove singularity originates from the valleys of the global minimum formed by the $\varDelta$~line:
\begin{equation}\label{eq:SC:DOS_at_tau_ast}
\rho_{\rm sc}(-3 + \delta\epsilon, \tau_\ast^{\rm sc}) = {A_{\rm sc}}/{\sqrt[4]{\delta\epsilon}} + o(1),
\end{equation}
where $A_{\rm sc} = \frac{\sqrt{2}}{\pi}{\Gamma(5/4)}/{\Gamma^3(3/4)} = 0.222$.

For the bcc lattice, the topology of the electron spectrum changes at $\tau^{\rm bcc}_{\ast} = 1$ (see~Fig.~
\ref{fig:ek}b and Table~\ref{table:bcc} in~the~Appendix). 
At $0 < \tau < \tau^{\rm bcc}_{\ast}$, there are two $\k$ points $\Gamma$ and P corresponding to local minima, one H~point corresponding to~a~maximum, and two saddle points $\varLambda^\ast/+1/$ and N$/-1/$. At $\tau = 0$,  the $\varLambda^\ast$ point coincides with P~point whereas the $w^{\rm bcc}_{\rm N}$, $w^{\rm bcc}_{\rm P}$ levels become merged; the line of van Hove $\k$ points $D = \{(\pi/2,\pi/2,k_z), 0<k_z<\pi/2\}$ is formed. This line generates a DOS singularity in the center of the band, $\rho_{\rm bcc}(\epsilon, \tau = 0) = (4/\pi^3)[\ln^2(64/|\epsilon|) - \pi^2/16] + o(\epsilon)$~\cite{1971:Katsura}. This relation is caused by the linear vanishing of inverse transverse masses $m^{-1}_D(k_z) = \pm 8\cos k_z$ at~the~P point on~the~$D$~line. If $\tau$ deviates from zero, the van Hove line $D$ is transformed to~a~van~Hove structure formed by three points: P (minimum), $\varLambda^\ast/+1/$, N$/-1/$, where $w^{\rm sc}_{\varLambda^\ast} - w^{\rm sc}_{\rm P} \ll w^{\rm sc}_{\rm N} - w^{\rm sc}_{\varLambda^\ast}$. 
In the limit $\tau\rightarrow 0$, all three masses diverge for the $\varLambda^\ast$ and P~points, whereas only one mass diverges at the N point. Within the narrow range $(w^{\rm sc}_{\rm P}$, $w^{\rm sc}_{\varLambda^\ast})$ , the DOS decreases steeply and an asymmetric plateau in the DOS is formed in the range $(w^{\rm sc}_{\varLambda^\ast}, w^{\rm sc}_{\rm N})$, where the DOS at $\epsilon = w^{\rm sc}_{\varLambda^\ast}$ is~much larger.

With an increase in $\tau$, the van Hove $\k$ point $\varLambda^\ast$ is shifted from the P~point to the $\Gamma$ point along the $\varLambda$~line. 
At $\tau > 2/3$, the roles of the P and $\Gamma$ points are interchanged. 
At $\tau = \tau^{\rm bcc}_{\ast}$, the $w^{\rm bcc}_{\rm N}$ and $w^{\rm bcc}_{\Gamma}$ levels merge together, giving rise to the $\varSigma$ line formed by van Hove $\k$~points. 
At $\tau \lesssim \tau^{\rm bcc}_{\ast}$, a three-point structure is formed, which is completely similar to that described above with the substitution P$\rightarrow\Gamma$ (see~Fig.~\ref{fig:w}b). 
At $\tau > \tau^{\rm bcc}_{\ast}$, the signatures of all three points change their signs and the role of the $\varLambda^\ast$ point is now played by the $\varDelta^\ast$~point, which is separated from the $\Gamma$ point at $\tau = \tau^{\rm bcc}_{\ast}$. At the further increase in $\tau$, this point migrates along the $\varDelta$ direction away from the $\Gamma$~point. 
In this case, the energy levels exhibit a behavior similar to that corresponding to the cases $0<\tau \ll \tau^{\rm bcc}_{\ast}$ and $\tau \lesssim \tau^{\rm bcc}_{\ast}$ discussed above. 
Similar to the situation with the $\varLambda^\ast$ point, at $\tau < \tau^{\rm bcc}_{\ast}$, all three masses diverge at $\tau\rightarrow\tau^{\rm bcc}_{\ast}$ also for the $\varDelta^\ast$ point. 
This means that an asymmetric plateau is formed on the $\rho_{\rm bcc}(\epsilon,\tau)$ plot at $\tau > \tau^{\rm bcc}_{\ast}$ between energies $w^{\rm bcc}_{\rm N}$ and $w^{\rm bcc}_{\varDelta^\ast}$, where $\rho_{\rm bcc}(\epsilon, \tau)$ is~much larger than $\rho_{\rm bcc}(w^{\rm bcc}_{\rm N},\tau)$, and the DOS decreases drastically between $w^{\rm bcc}_{\varDelta^\ast}$ and $w^{\rm bcc}_{\Gamma}$ energy levels. 
The maximum DOS value for $\tau < \tau^{\rm bcc}_{\ast}$ and $\tau > \tau^{\rm bcc}_{\ast}$ is achieved at $\epsilon = w^{\rm bcc}_{\varLambda^\ast}$ and $\epsilon = w^{\rm bcc}_{\varDelta^\ast}$, respectively, which correspond to internal van Hove $\k$ points in the Brillouin zone (see Fig.~\ref{fig:DOS_vs_tau}b in~the~Appendix). 
Thus, the van Hove lines $D$ and $\varSigma$ for the bcc lattice are always transformed to a three-point van Hove structure including the broad asymmetric plateau and the adjacent interval with a steep decrease in the DOS. 
Some electron spectra calculated for different $\tau$ values in~symmetric directions of~the~Brillouin zone for~the~sc and bcc lattices are~given in Fig.~\ref{fig:ek} in the Appendix.  

At $\tau = \tau_\ast^{\rm bcc}$, the DOS has the asymptotic behavior
\begin{equation}
	\rho_{\rm bcc}(-2 + \delta\epsilon, 1) =
	-\frac{5}{2\pi^2} +
	\begin{cases}
		{B_{\rm bcc}}/{\sqrt[4]{-\delta\epsilon}},& \delta\epsilon < 0\\
		{A_{\rm bcc}}/{\sqrt[4]{\delta\epsilon}},& \delta\epsilon > 0
	\end{cases}  + o(1),
\end{equation}
where
$
A_{\rm bcc} = \frac3{2\pi^{3}}
	\int\limits_{+1}^{+\infty}
	{dt\mathbb{K}(1 - t^2)}/{\sqrt{t}} = 0.333,\; B_{\rm bcc} = \frac1{\pi^3}\int\limits_0^{+\infty}{dt}\mathbb{K}(-t^2)/{\sqrt{t}} = 0.314$, $\mathbb{K}$ being complete elliptic integral of the first kind~(see the~Appendix).

The direct study of the van Hove $\k$ points for the fcc lattice is illustrated in Table~\ref{table:fcc} in the Appendix. In~Fig.~\ref{fig:w_fcc}, we can see that the energy surfaces in the spectrum undergo topological transformations (changes in the type of van Hove points) at $\tau = -1, -1/2,0,1/2,1$. Here, a whole van Hove surface given by the equation $\cos k_xx + \cos k_y + \cos k_z = 0$ is formed at $\tau = -1/2$. The energy level for this surface corresponds to the merged levels for the W and L points. At $\tau = 0$, the van Hove line $V$ is formed, which corresponds to the merged levels for the W and~X~points and connects these points. 

\begin{figure}[t!]
\noindent
\includegraphics[width=0.45\textwidth]{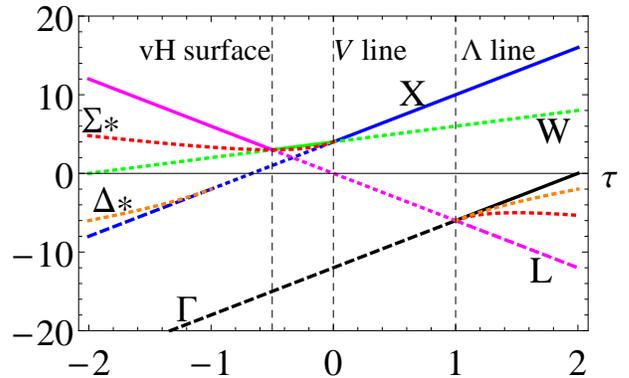}
\caption{(Color online) $\tau$ dependence of the van Hove levels $w$  for the fcc lattice. The notation is the same as in Fig.~\ref{fig:w}.
\label{fig:w_fcc}
}
\end{figure}

At $\tau = 1$, the van Hove line~$\varLambda$ is~formed, which corresponds to the merged levels for the $\Gamma$ and L~points and connects these points. 
One or several masses diverge at the singular points $\tau = -1/2, 0, 1$. This means that the deviation of $\tau$ from the ``critical'' value is accompanied by the splitting of the van Hove surface or line into several van Hove points corresponding to a large mass. 
In particular, two van Hove points W$/+1/$ and $\varSigma^\ast/-1/$ with~the~closely located energy levels ($w^{\rm fcc}_{\varSigma^\ast}(\tau) - w^{\rm fcc}_{\rm W}(\tau)\sim \tau + 1/2$) are formed at~$\tau \lesssim -1/2$. Therefore, a narrow DOS plateau arises between these levels.

At $-1/2<\tau<0$, $\k$ points form a van Hove structure corresponding to the narrow energy range $(w^{\rm fcc}_{\varSigma^\ast}, w^{\rm fcc}_{\rm W})$ (its~width does not exceed 0.35). 
A pair of such levels gives rise to a pronounced drop in the DOS from the maximum to the saddle point characterized by large masses. 
Note that according to Table~\ref{table:fcc} in~the~Appendix, the internal van Hove points $\varDelta^\ast$ and $\varSigma^\ast$ exist at $|\tau| > 1$ and |$|\tau - 1/2| > 1/2$, respectively. 
The least interesting case $0 < \tau < 1$  was analyzed in detail in~Ref.~\cite{1972:Swendsen}. 
In this case, there are neither internal van Hove points nor van Hove lines and surfaces. 
At $\tau > 1$, the van Hove line $\varLambda$, existing at $\tau = 1$, is split into three van Hove points $\varDelta^\ast$, $\varSigma^\ast$, $\Gamma$. 
In this case, the energy levels of the first two points form a typical narrow plateau ($w^{\rm fcc}_{\varSigma^\ast}(\tau) - w^{\rm fcc}_{\varDelta^\ast}(\tau)\sim (\tau - 1)^2$) between the levels of saddle points with opposite signatures. This situation is completely similar to the splitting of the van Hove line $\varSigma$ for the sc lattice, which produces a narrow quasisymmetric plateau and an interval of sharp decrease. Thus, we can make an a priori conclusion about the existence of a narrow stable plateau in the DOS plot at $\tau\gtrsim+1$.

Let us analyze the types of the DOS singularities related to the found van Hove surface and lines. Since $t^{\rm fcc}_\mathbf{k}(\tau = -1/2) = 3 - (1/2)\left(t^{\rm sc}_\mathbf{k}(\tau = 0)\right)^2$~\cite{1998:Ulmke}, then
\begin{equation}
	\rho_{\rm fcc}(\epsilon; \tau = -1/2) = \sqrt{\frac2{3 - \epsilon}}\rho_{\rm sc}(\sqrt{2(3 - \epsilon)}; \tau = 0).
\end{equation}
Thus, the van Hove surface generates a giant ``one-dimensional'' singularity at the band top $\rho_{\rm fcc}(\epsilon)\sim (3 - \epsilon)^{-1/2}$. 

The inverse transverse masses for the van Hove line $V = \{(0,\pi,k_z), -\pi/2<k_z<\pi/2\}$ at $\tau = 0$ are $m^{-1}_{V}(k_z) = -4(1 \pm \cos k_z)$. 
Only one of these inverse masses vanishes quadratically; therefore, the geometric mean mass vanishes linearly. This means that the van Hove singularity arising near the $V$~line has the form
\begin{equation}
\rho_{\rm fcc}(\epsilon; \tau = 0) \simeq (3/8\pi^2)\ln[8/(4-\epsilon)].
\end{equation}
The coinciding inverse transverse masses for the line $\varLambda = \{k_x = k_y = k_z = k_\lambda, -\pi/2<k_\lambda<\pi/2\}$ at~$\tau = 1$ have the single value $m^{-1}_{\varLambda}(k_\Lambda) = 12\sin^2k_\Lambda$ vanishing quadratically at $k_\Lambda = 0$; therefore, the geometric mean mass also vanishes quadratically. This gives $\rho_{\rm fcc}(\epsilon; \tau = +1)\sim [\theta(\epsilon-6)(\epsilon-6)]^{-1/4}$ and generates a new giant van Hove singularity ($\theta(x)$ is~the~Heaviside step function). 

\section{Electronic properties}
\begin{figure*}
\includegraphics[angle=-90,width=0.45\textwidth]{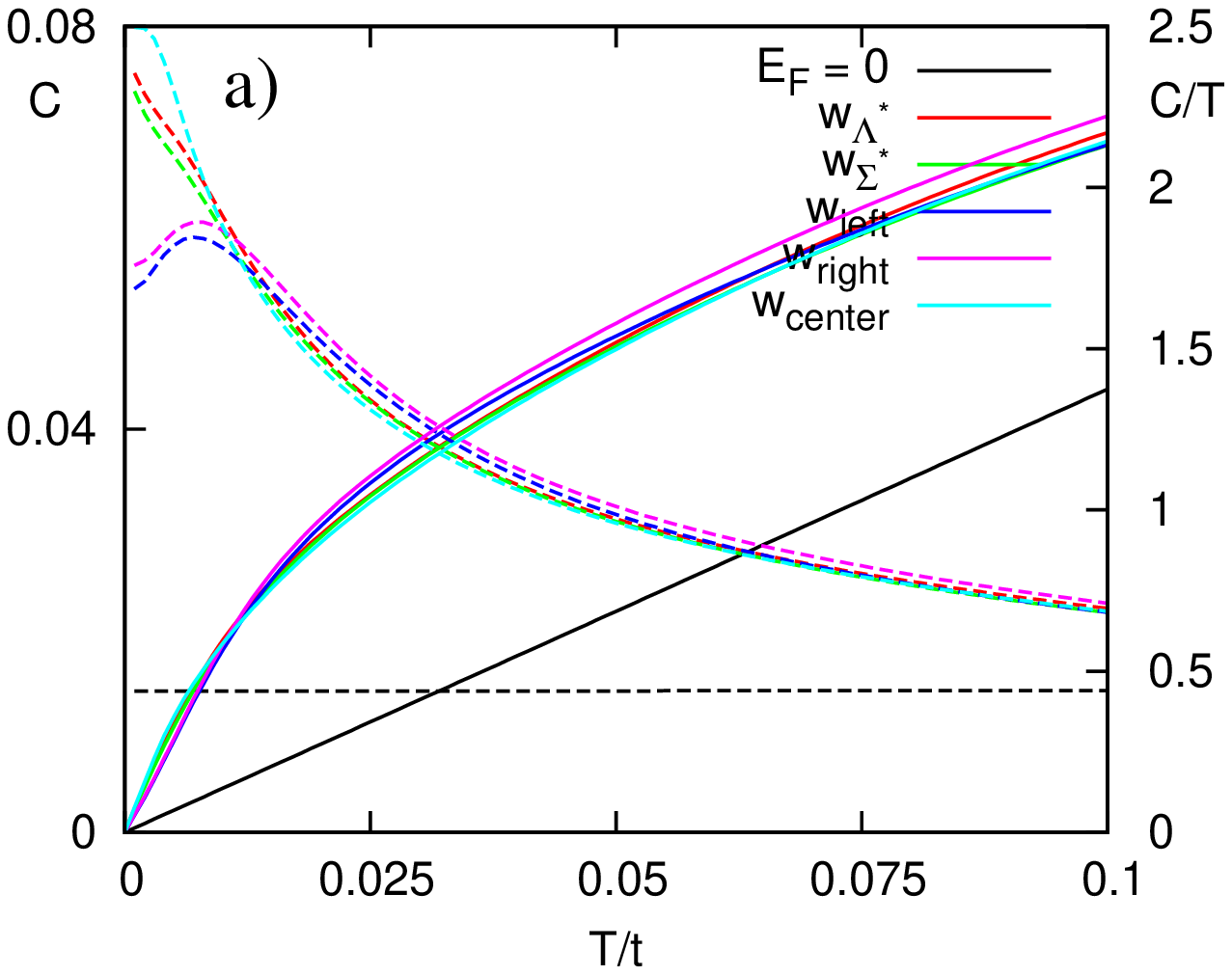}
\includegraphics[angle=-90,width=0.45\textwidth]{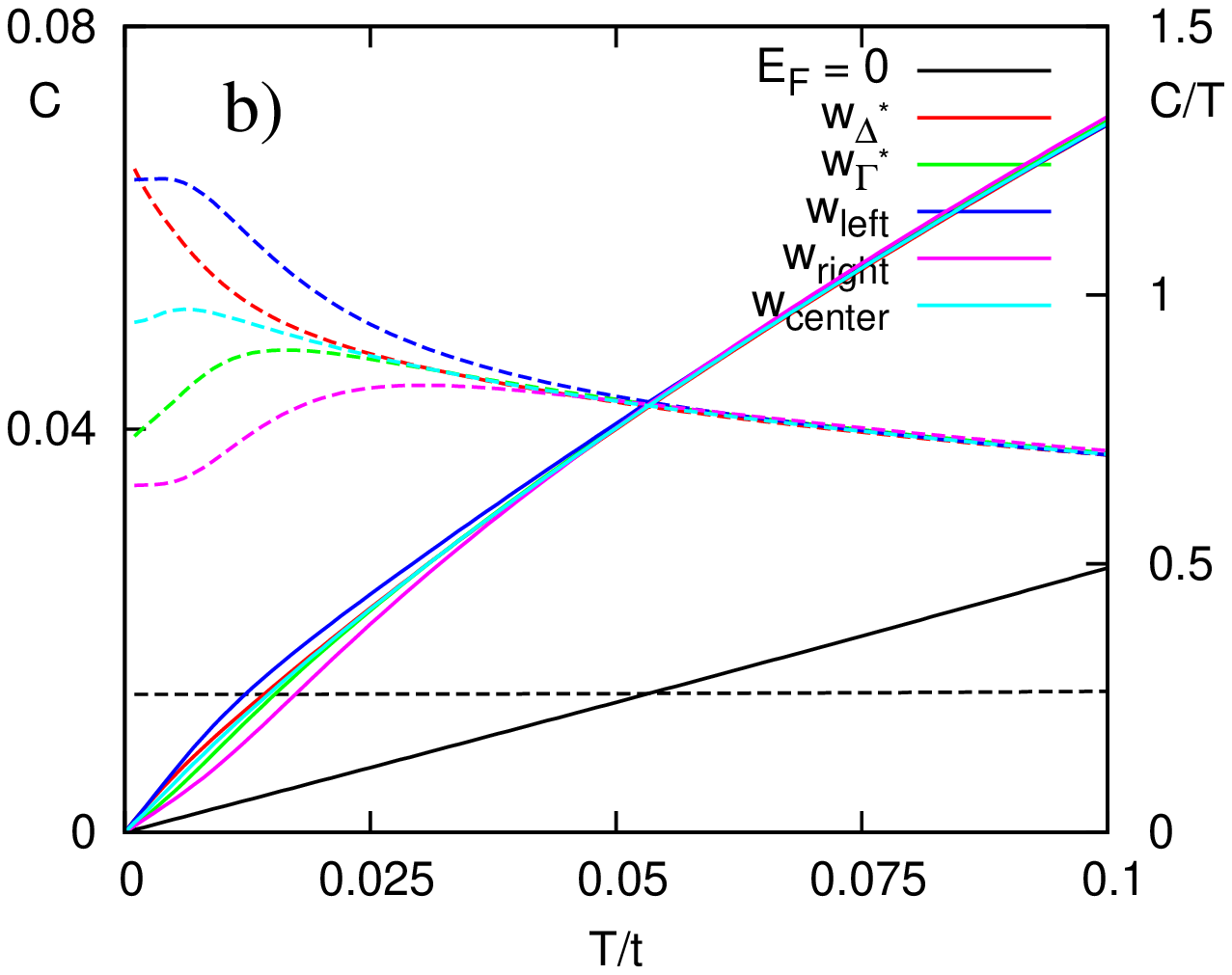}
\caption{
Temperature dependence of (left axis, solid lines) the specific heat $C$ and of (right axis, dashed lines) coefficient $\gamma = C/T$ at different positions of the Fermi level for (a) the sc lattice ($\tau = 0.3$): 
$w^{\rm sc}_{\varSigma^\ast} = -2.533, w^{\rm sc}_{\varLambda^\ast} = -2.500, w^{\rm sc}_{\rm center} = (1/2)(w^{\rm sc}_{\varSigma^\ast} + w^{\rm sc}_{\varLambda^\ast}) = -2.517, w^{\rm sc}_{\rm left} = -2.550, w^{\rm sc}_{\rm right} = -2.480$. (b) bcc lattice~($\tau = 1.1$): $w^{\rm bcc}_{\varDelta^\ast} = -1.436, w^{\rm bcc}_{\Gamma} = -1.400, w^{\rm bcc}_{\rm center} = -1.418, w^{\rm bcc}_{\rm left} = -1.470, w^{\rm bcc}_{\rm right} = -1.370$.
\label{fig:C_vs_T}
}
\end{figure*}

In Fig.~\ref{fig:C_vs_T}, we show the temperature dependence of the electronic specific heat 
$
	C(T; E_{\rm F}) = -\frac1{T}\int d\epsilon \rho(\epsilon)(\epsilon - E_{\rm F})^2f'(\epsilon),
$	
where $f(E) = (\exp[(E - E_{\rm F})/T] + 1)^{-1}$ is the Fermi function for~sc~and~bcc lattices at low temperatures obtained in the free-electron approximation using the exact expression for the density of states presented in~the~Appendix. 
The results are shown for the $\tau$ values close to the corresponding topological transitions and the Fermi level $E_{\rm F}$ located inside or near the levels of the van Hove structure, which remains after the splitting of the van Hove lines $\varDelta$ and $\varSigma$ for the sc and bcc lattices, respectively~(see discussion above).
For clarity, we also show the results of calculation for the case $E_{\rm F} = 0$, when the Fermi level lies far away from all singularities. In the same approximation, we calculate the magnetic susceptibility (in units of $\mu_{\rm B}^2$)
$
	\chi(T; E_{\rm F}) = -\int d\epsilon \rho(\epsilon)f'(\epsilon)
$
and thermopower
$
	S(T; E_{\rm F}) = -[1/T\sigma(T; E_{\rm F})]\int d\epsilon (\epsilon - E_{\rm F})\sigma_{\rm c}(\epsilon)f'(\epsilon),
$
where the electrical conductivity is defined as
$
	\sigma(T; E_{\rm F}) = -\int d\epsilon \sigma_{\rm c}(\epsilon)f'(\epsilon),
$
with the conductivity function $\sigma_{\rm c}(\epsilon)$ 
proportional to the density of states~\cite{1993:Trefilov}.

\begin{figure*}
\includegraphics[angle=-90,width=0.45\textwidth]{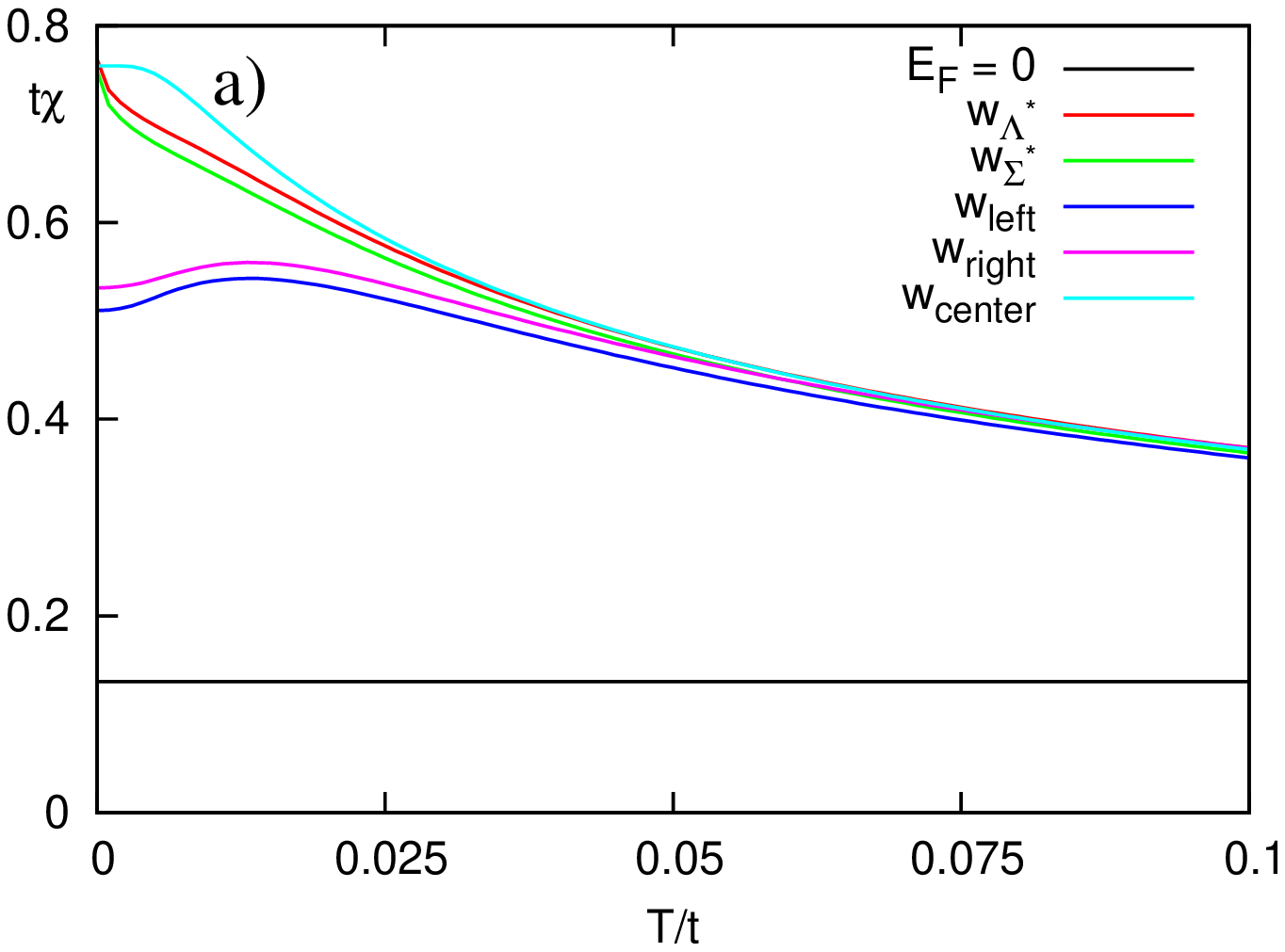}
\includegraphics[angle=-90,width=0.45\textwidth]{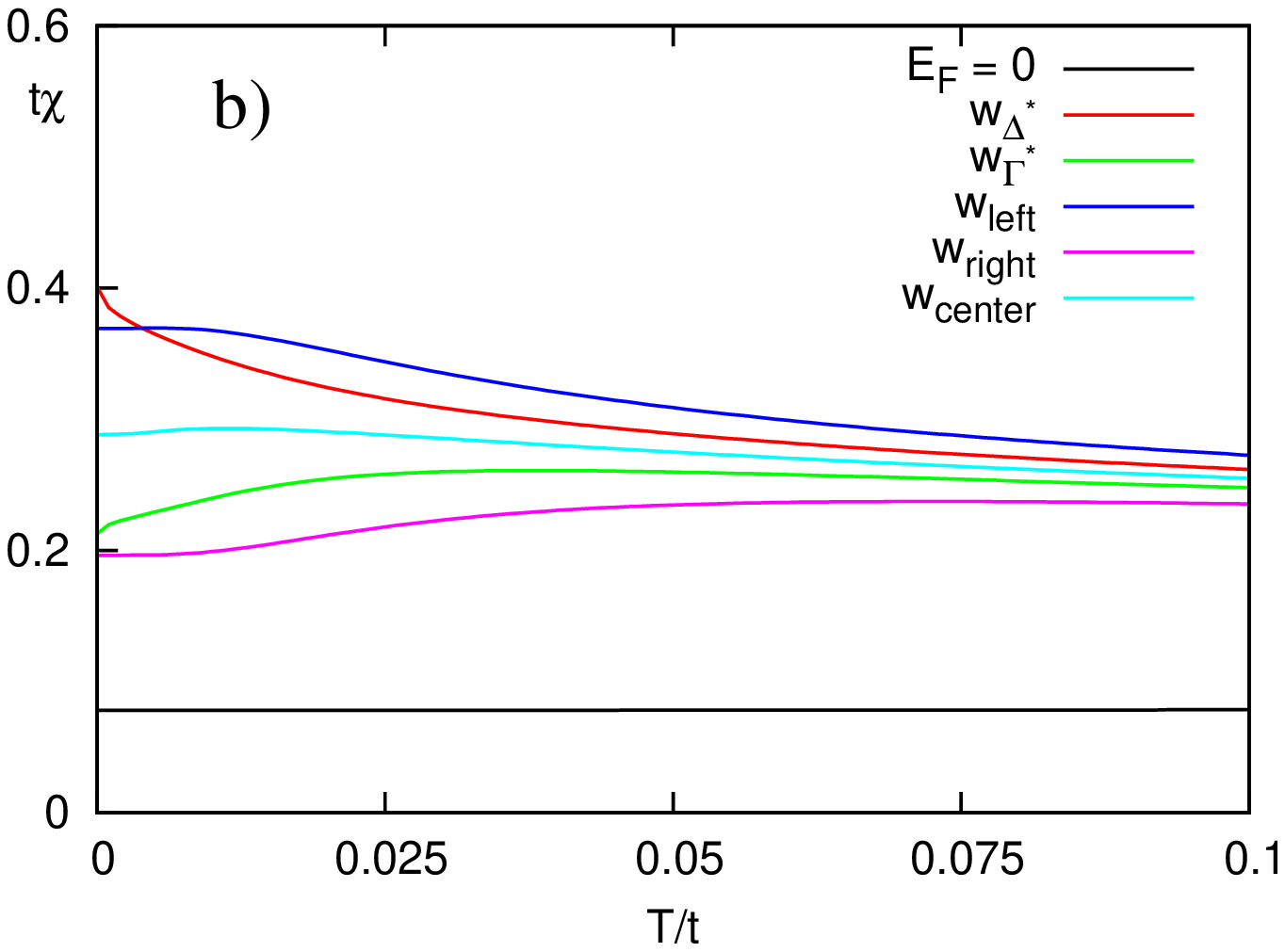}
\caption{Temperature dependence of the magnetic susceptibility $\chi$ at different positions of the Fermi level for the
(a) sc lattice ($\tau = 0.3$) and (b) bcc lattice ($\tau = 1.1$).
\label{fig:chi_vs_T}
}
\end{figure*}
\begin{figure*}
\includegraphics[angle=-90,width=0.45\textwidth]{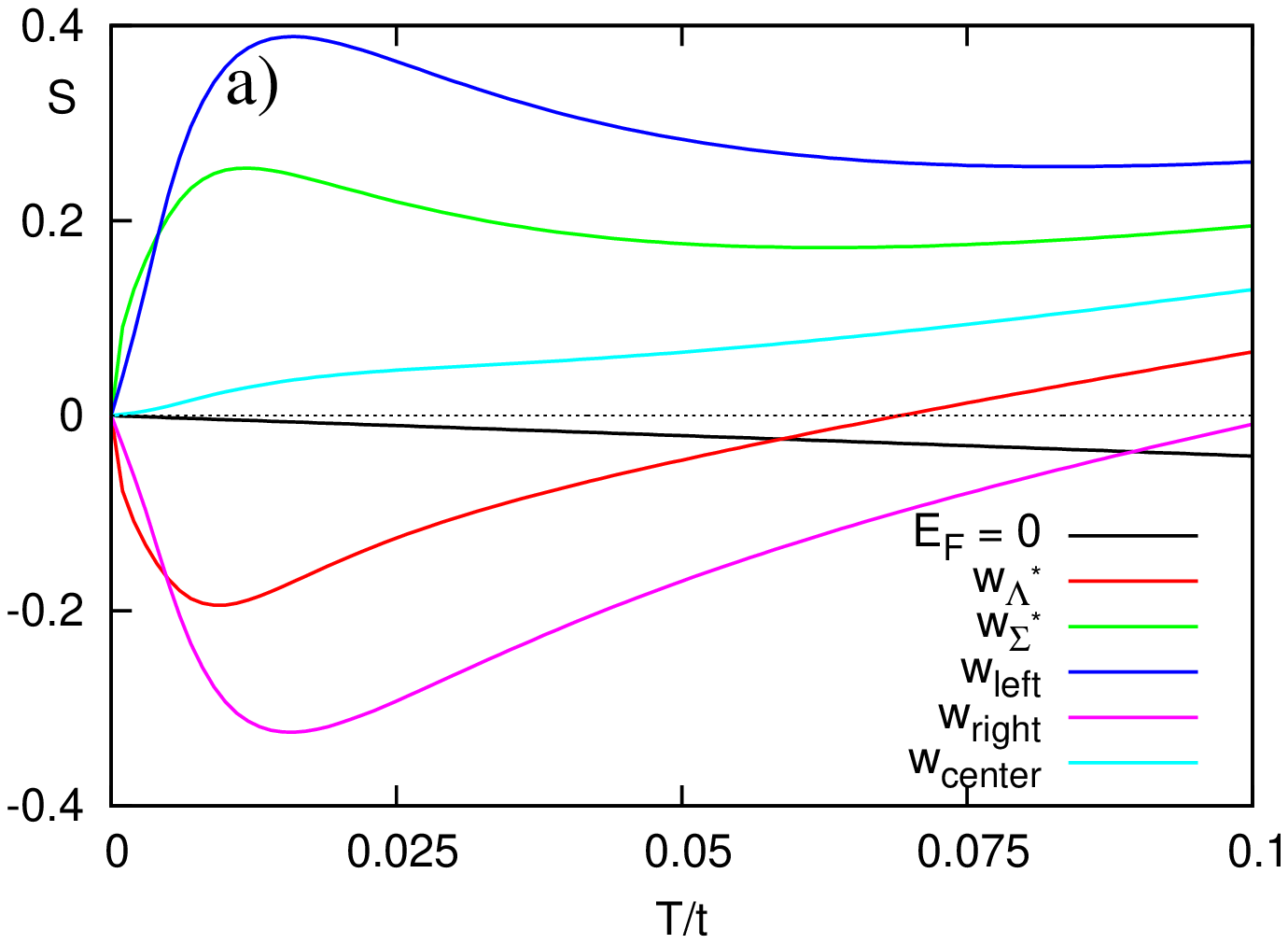}
\includegraphics[angle=-90,width=0.45\textwidth]{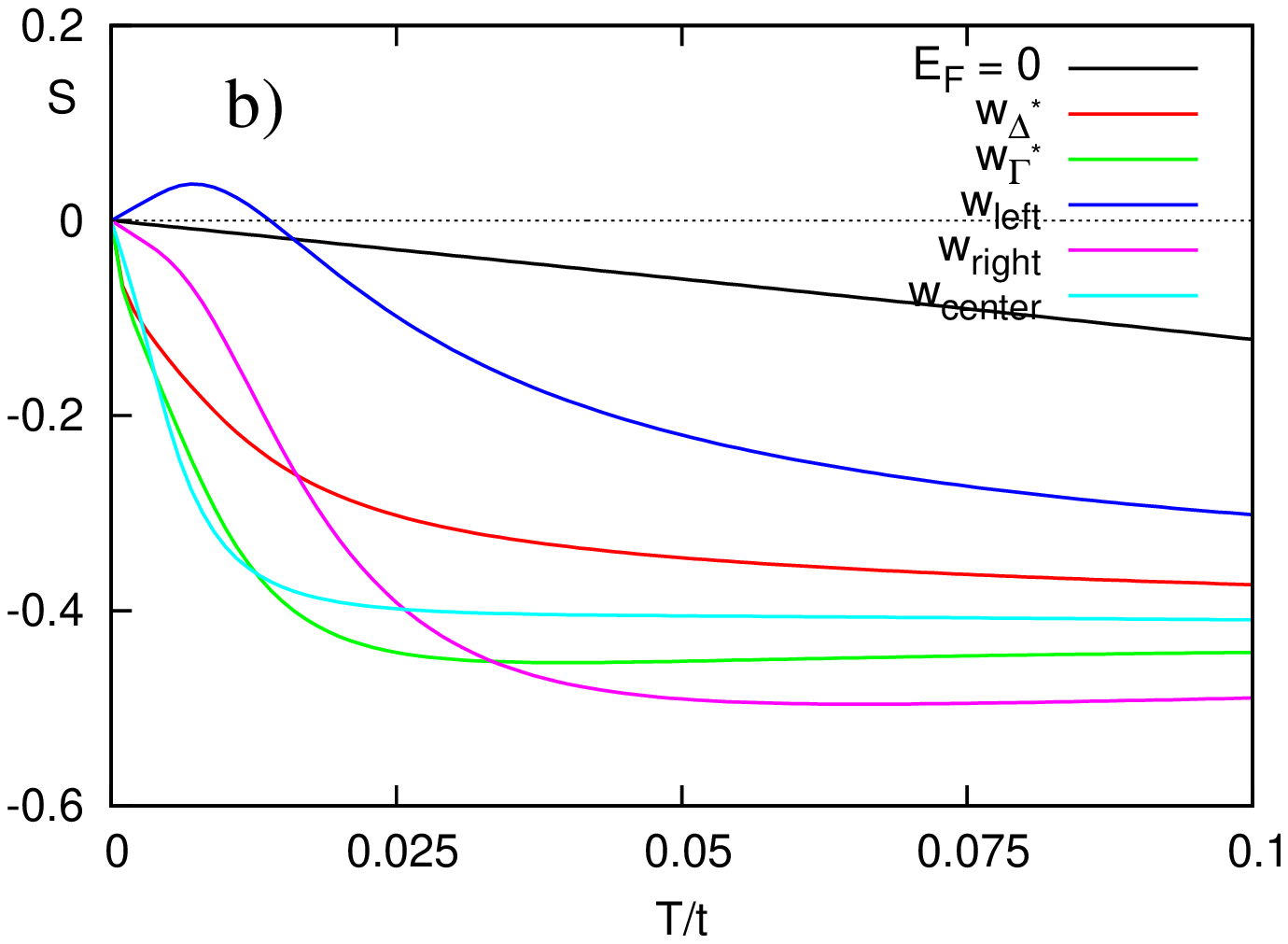}
\caption{Temperature dependence of the thermopower~$S$  at~different positions of the Fermi level for the (a) sc lattice ($\tau = 0.3$) and (b) bcc lattice ($\tau = 1.1$).
\label{fig:S_vs_T}
}
\end{figure*}

It is seen that the specific heat for the sc and bcc
lattices is high and strongly deviates from the standard
linear temperature dependence (at~$E_{\rm F} = 0$). 
This behavior remains the same far beyond the temperature
range of the order of the peak width, $T_\ast/t \approx 0.03$.
If we choose $t\sim0.5$~eV, then $T_\ast \sim 200$~K. The specific heat also strongly depends on the position of the Fermi level within the van Hove structure at~$T \ll T_\ast$. If $E_{\rm F}$ is located at the DOS plateau of~the~sc lattice (see Fig.~\ref{fig:DOS_vs_e}a in~the~Appendix), the~specific
heat will exhibit a linear behavior (with a giant slope $\gamma$) only within a~narrow low-temperature range; upon
deviation of $E_{\rm F}$ to the left or right of~the~plateau borders, $\gamma(T)$ plot acquires a~maximum.
For bcc lattice, the behavior $\gamma$ is affected by the existence of an asymmetric peak in the energy interval $(w^{\rm bcc}_{\varLambda^\ast}, w^{\rm bcc}_{\Gamma})$.
Indeed, when $E_{\rm F} = w^{\rm bcc}_{\varLambda^\ast}$ or lower, $\gamma(T)$ decreases monotonically with the growth
of~the~temperature; whereas $E_{\rm F} > w^{\rm bcc}_{\varLambda^\ast}$, $\gamma(T)$  exhibits
a low-temperature peak. At $T \gtrsim T_\ast$   the nonlinear behavior of~the~specific heat still takes place, although the dependence on the position of the Fermi level
almost disappears. 

The magnetic susceptibility $\chi$ (Fig.~\ref{fig:chi_vs_T}) exhibits a non-Pauli behavior; namely, it strongly depends on the temperature. For the sc lattice, the narrow DOS plateau favors the localization of electrons, so that the behavior of the magnetic susceptibility is close to the Curie-Weiss law. If the Fermi level is located at the top of the plateau and $\tau$ is quite close to $\tau^{\rm sc}_\ast$, the DOS is high and nearly insensitive to changes in the Fermi level within the top of the plateau. This means that ferromagnetism will be quite stable also with respect to the correlation effects (higher Curie temperature), in contrast to ferromagnetism originating from the van Hove point corresponding to a large mass (see the discussion in the Introduction). The shift of the Fermi level outside the plateau leads to a peak in $\chi (T)$ and hence to the possible formation of temperature-induced ferromagnetism~\cite{Vons1}. For the bcc lattice, the temperature dependence of $\chi$ is not so strong, although its low-temperature value substantially
depends on $E_{\rm F}$.

The asymmetry of the density of states with respect to $E_{\rm F}$ leads to a large thermopower. For both lattices, it nonmonotonically depends on the temperature, and the character of this dependence strongly depends on the position of $E_{\rm F}$ (Fig.~\ref{fig:S_vs_T}). For the sc lattice, the temperature dependence is symmetric in magnitude and antisymmetric in sign with respect to the position of the Fermi level measured from the center of the
plateau.

\section{Conclusions}
Taking into account electron hoppings between the nearest and next-nearest neighbor sites, we have demonstrated that the energy surfaces located near the van Hove $\k$ point can undergo topological transitions accompanied by changes in the parameters of the electron spectrum. In the course of such a transition, the energy levels of van Hove points merge, forming van Hove lines (or surfaces), which give rise to giant (logarithmic or power law) singularities in the density of states. The presence of these singularities can lead to new instabilities (magnetic or superconducting) and to an unusual temperature dependence of the observed
physical characteristics. Near~the~topological transition, the van Hove lines and surfaces are transformed to three-point van Hove structures manifesting themselves either as a narrow plateau leading to the localization of a part of the electron states (then, the thermodynamic parameters can be represented as the sum of the contributions from the localized and itinerant electron states), with an adjacent wider interval characterized by a sharp decrease in the DOS, or as the range of a sharp decrease in the DOS with the adjacent asymmetric plateau (which leads to the intermediate ``semilocalized'' behavior of the corresponding thermodynamic contribution). 

The corresponding effects should be enhanced because of~correlation effects~\cite{1993:Trefilov}. Thus, the results obtained (as applied to the spectrum of both Fermi and Bose excitations) can be used to study magnetic and electronic characteristics in the framework of various microscopic many-electron models, for example, the Heisenberg, Kondo, and Hubbard models.

The relationship of the van Hove points corresponding to a large mass and the formation of ferromagnetism can be illustrated by a number of examples. Namely, ZrZn$_2$ is close to a Lifshitz transition, which manifests itself in a change in the topology of the Fermi surface with a shift in $E_{\rm F}$~\cite{2001:Santi}. The weak ferromagnetism of ZrZn$_2$ is probably caused by the contribution to the density of states from the van Hove points near the faces of the Brillouin zone (X, L, and K points). Large, but finite, values of the corresponding effective masses obtained by \textit{ab initio} calculations~\cite{2001:Santi} cut off the singularity of the density of states near $E_{\rm F}$.

The large mass at the van Hove $\mathbf{k}$ point R provides weak ferromagnetism (with a Curie temperature of
41~K) in~Ni$_3$Al~\cite{2011:Hamid}. The itinerant weak ferromagnet Y$_2$Ni$_7$ exhibits a~low Curie temperature of 58~K and a~low saturation magnetic moment of $0.44~\mu_{\rm B}$~\cite{Y2Ni7:Inoue,Y2Ni7:Nishihara}.
The band structure calculations yield the spectrum
exhibiting flat bands near the van Hove $\Gamma$ point and a~dispersionless spectrum in the P--Z direction of the
rhombohedral Brillouin zone (a possible van Hove
line or a pair of van Hove points with a large mass)~\cite{2015:Singh}.

The performed analysis can be used to supplement computational methods used to calculate the electron
spectrum and the properties of actual materials. In particular, the effect of van Hove points on the electron density of states cannot be satisfactorily explored in the framework of the tetrahedron method~\cite{1994:Andersen}. As an alternative, we can suggest the analytical approximation of the spectrum near high-symmetry points, where the spectrum is almost flat (nearly dispersionless)~\cite{2017:Stepanenko}.

\section{Acknowledgments}
We are grateful to M.~I.~Katsnelson, A.~O.~Anokhin, and~A.~A.~Katanin for~valuable discussions. 

This work was supported by the Ministry of Science and
Higher Education of the Russian Federation (state assignment no. AAAA-A18-118020190095-4, project Quantum)
and by the Government of the Russian Federation (program 211, state contract no. 02.A03.21.0006).

\appendix
\counterwithin{figure}{section}
\section{Supplemental Material}
\label{sec:appendix}
\begin{table*}[t!]
\caption{\label{table:sc} Non-equivalent $\k$ points of van Hove singularities for sc lattice, see.~Fig.~\ref{fig:w}a of main text, $\tau \ge 0$. Arrow denotes a change of signature of mass tensor of van Hove $\mathbf{k}$ point as  $\tau$ increases above these value $\tau^{\rm sc}_\ast=1/4$. $k^{\rm sc}_{\varSigma^\ast} = \arccos\left[(2\tau)^{-1} - 1\right], k^{\rm sc}_{\varLambda^\ast} = \arccos\left[(4\tau)^{-1}\right]$}
\center
\begin{tabular}{|c|c|c|c|}
\hline
$\mathbf{k}$ & $w^{\rm sc} = t(\mathbf{k})$ & \mbox{inverse masses} & \mbox{signature} \\
\hline
$\Gamma(0,0,0)$ & $ -6 + 12\tau$ & $2(1-4\tau),2(1-4\tau), 2(1-4\tau)$&min$\stackrel{\tau = 1/4}{\rightarrow}$max\\
\hline
R$(\pi,\pi,\pi)$ & $ +6 + 12\tau$ & $-2(1+4\tau),-2(1+4\tau),-2(1+4\tau)$&\\
\hline
X$(0,0,\pi)$ & $ -2 - 4\tau$ &$2,2,2(4\tau - 1)$& $/+1/\stackrel{\tau = 1/4}{\rightarrow}$min \\
\hline
M$(0,\pi,\pi)$ & $ +2 - 4\tau$ & $-2,-2,2(4\tau + 1);$&  $/-1/$\\
\hline
$\varSigma^\ast(0, k^{\rm sc}_{\varSigma^\ast}, k^{\rm sc}_{\varSigma^\ast})$ & $ -\tau^{-1} + 2 - 4\tau$ & $2(4\tau - 1), 4 - \tau^{-1}, \tau^{-1} - 4$ &$\tau > 1/4$,$/+1/$  \\
\hline
$\varLambda^\ast(k^{\rm sc}_{\varLambda^\ast}, k^{\rm sc}_{\varLambda^\ast}, k^{\rm sc}_{\varLambda^\ast})$ & $ -3\tau^{-1}/4$& $\tau^{-1}((4\tau)^2 - 1),(2\tau)^{-1}(1 - (4\tau)^2),(2\tau)^{-1}(1 - (4\tau)^2)$ &$\tau > 1/4$, $/-1/$ \\
\hline
\end{tabular}
\end{table*}

Formal definition of density of state reads
\begin{equation}\label{eq:DOS_def}
	\rho(\epsilon; \tau) = \frac1{N}\sum_{\mathbf{k}}\delta(\epsilon - t(\mathbf{k}, \tau)),
\end{equation}
where $t(\mathbf{k}, \tau)$ is the spectrum within tight-binding approximation. 
We choose the signs of transfer integrals in the spectrum as follows
\begin{equation*}
\begin{split}
t_{\rm sc}(\mathbf{k}, \tau) &= -2(\cos k_x + \cos k_y + \cos k_z) +{}\\ &+4\tau (\cos k_y\cos k_z + \cos k_z\cos k_x + \cos k_x\cos k_y),
\end{split}
\end{equation*}
\begin{equation*}
\begin{split}
	t_{\rm bcc}(\mathbf{k}, \tau) &= -8\cos k_x\cos k_y\cos k_z +{}\\ &+2\tau (\cos 2k_x + \cos 2k_y + \cos 2k_z),
\end{split}
\end{equation*}
\begin{equation*}
\begin{split}
	t_{\rm fcc}(\mathbf{k}, \tau) &= -4(\cos k_y\cos k_z + \cos k_x\cos k_z  + \cos k_x\cos k_y) \\
	&+ 2\tau(\cos 2k_x + \cos 2k_y + \cos 2k_z),
\end{split}
\end{equation*}
where lattice constant is taken as unity. 
For the van Hove point $\bar{\mathbf{k}}$ we expand the spectrum
\begin{equation}\label{eq:spectrum_exp}
	t(\mathbf{k}) = t(\bar{\mathbf{k}}) + \frac12\sum_{ij}\frac{\partial^2t(\bar{\mathbf{k}})}{\partial k_i\partial k_j}(k_i - \bar{k}_{i})(k_j - \bar{k}_{j}).
\end{equation}
Let $a_i$ be the eigenvalues of the matrix ${\partial^2t(\bar{\mathbf{k}})}/{\partial k_i\partial k_j}$~(inverse mass tensor). We introduce the signature as the difference between numbers of positive and negative 
eigenvalues of mass tensor  
and write it in slashes, e.g.,~maximum (minimum) corresponds to $/+3/(/-3/)$, saddle points to $/\pm1/$.
For three-dimensional lattice in the non-degenerate case a local minimum~(maximum) of the spectrum $t(\mathbf{k})$ corresponds to one-side square-root increase~(decrease) of DOS as the energy $\epsilon$ deviates from $\epsilon_0$:
\begin{equation}
	\rho(\epsilon) = \rho(\epsilon_0) + A\sqrt{\theta(\pm(\epsilon - \epsilon_0))|\epsilon - \epsilon_0|} + O(\epsilon - \epsilon_0)
\end{equation}
in the vicinity of the van Hove level $\epsilon_0$; a saddle-type van Hove point with the  mass tensor signature $/+1/$~($/-1/$) corresponds to one-side square-root decreasing,
\begin{equation}
	\rho(\epsilon) = \rho(\epsilon_0) - A\sqrt{\theta(\mp(\epsilon - \epsilon_0))|\epsilon - \epsilon_0|} + O(\epsilon - \epsilon_0)
\end{equation}
with positive constant $A = 2\pi V_{\rm BZ}|a_1a_2a_3|^{-3/2}$, $V_{\rm BZ}$ being the Brillouin zone volume.

\begin{figure*}
\centering
\includegraphics[angle=-90,width=0.9\textwidth]{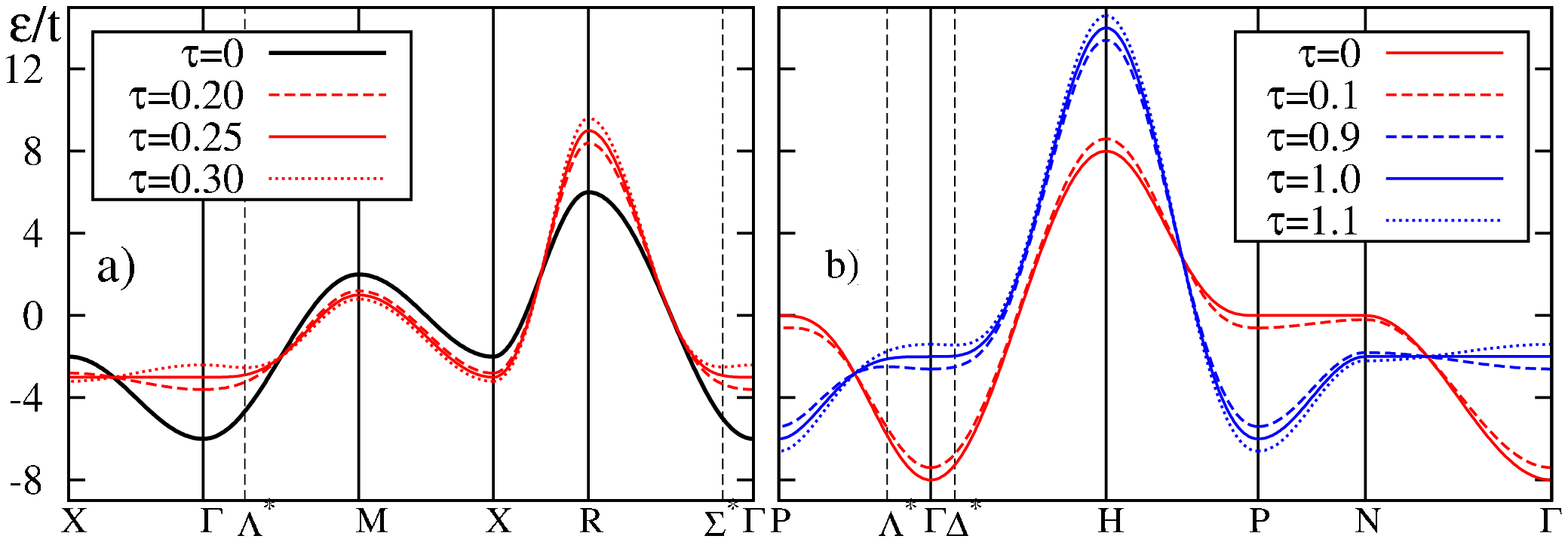}
\caption{Energy spectrum in high-symmetry directions of the Brillouin zone (``spaghetti''). (a) sc, (b) bcc lattice, van Hove $\mathbf{k}$ points are shown.
\label{fig:ek}
}
\end{figure*}

Fig.~\ref{fig:ek} shows the spectrum for sc and bcc lattice in high-symmetry directions of the Brillouin zone for $\tau$ being in the vicinity of topological transition.

\begin{table*}[t!]
\caption{\label{table:bcc} Non-equivalent $\k$-points of van Hove singularities for bcc lattice, see.~Fig.~\ref{fig:w}b of main text, $\tau \ge 0$. The notations are the same as in the Table~1. $k^{\rm bcc}_{\varLambda^\ast} = \arccos\tau, k^{\rm bcc}_{\varDelta^\ast} = \arccos\tau^{-1}$.}
\center
\begin{tabular}{|c|c|c|c|}
\hline
$\mathbf{k}$ & $w^{\rm bcc} = t(\mathbf{k})$ & \mbox{inverse masses} & \mbox{signature} \\
\hline
$\Gamma(0,0,0)$ & $-8 + 6\tau$& $8(1-\tau),8(1-\tau),8(1-\tau)$& min$\stackrel{\tau = 1}{\rightarrow}$max\\
\hline
H$(0,0,\pi)$ & $+8 + 6\tau$ & $-8(1+\tau),-8(1+\tau),-8(1+\tau)$&max\\
\hline
P$(\pi/2,\pi/2,\pi/2)$ & $-6\tau$ &$8\tau,8\tau,8\tau$& min \\
\hline
N$(\pi/2,\pi/2,0)$ & $-2\tau$ &$-4\tau,4(1 + \tau),4(-1 + \tau)$& $/-1/\stackrel{\tau = 1}{\rightarrow}/+1/$ \\
\hline
$\varLambda^\ast\left(k^{\rm bcc}_{\varLambda^\ast}, k^{\rm bcc}_{\varLambda^\ast}, k^{\rm bcc}_{\varLambda^\ast}\right)$ & $2\tau(2\tau^2 - 3)$ & $16\tau(1-\tau^2),16\tau(1-\tau^2),-8\tau(1-\tau^2)$& $\tau < 1$, $/+1/$\\
\hline
$\varDelta^\ast\left(0, 0, k^{\rm bcc}_{\varDelta^\ast}\right)$ & $2\tau - 4\tau^{-1}$& $8(\tau^{-1} - \tau),8(\tau^{-1} - \tau),8(\tau - \tau^{-1})$ &$\tau > 1$, $/-1/$\\
\hline
\end{tabular}
\end{table*}
\begin{table*}[t!]
\caption{\label{table:fcc} Non-equivalent $\k$ points of van Hove singularities for FCC lattice, see.~Fig.~\ref{fig:w_fcc} of main text. The notations are the same as in the Table~\ref{table:sc}. $k^{\rm fcc}_{\varSigma^\ast} =  \arccos(2\tau - 1)^{-1}, k^{\rm fcc}_{\varDelta^\ast} = \arccos\tau^{-1}$. $a^{\rm fcc}_{\rm \varDelta^\ast1} = 4(\tau^{-1}-1)(1 + 2\tau),
a^{\rm fcc}_{\rm \varDelta^\ast2} = 8\tau^{-1}(\tau^2 - 1),
a^{\rm fcc}_{\rm \varSigma^\ast1} = 8(\tau - 1)(1 + 2\tau)/(1 - 2\tau),	
a^{\rm fcc}_{\rm \varSigma^\ast2} = 16\tau(\tau - 1)(1 + 2\tau)(1 - 2\tau)^{-2},
a^{\rm fcc}_{\rm \varSigma^\ast3} = 16\tau(\tau - 1)(2\tau - 1)^{-1}
$.}
\center
\begin{tabular}{|c|c|c|c|}
\hline
$\mathbf{k}$ & $w^{\rm fcc} = t(\mathbf{k})$ & inverse masses & signature \\
\hline
$\Gamma(0,0,0)$ & $-12 + 6\tau$ & $8(1 - \tau),8(1 - \tau),8(1 - \tau)$&min$\stackrel{\tau = 1}{\rightarrow}$max\\
\hline
X$(0,0,\pi)$ & $+4 + 6\tau$ & $-8\tau,-8\tau, -8(1+\tau)$&min$\stackrel{\tau = -1}{\rightarrow}/+1/\stackrel{\tau = 0}{\rightarrow}{\rm max}$\\
\hline
W$(0,\pi/2,\pi)$ & $+4 + 2\tau$ &$-4(1 + 2\tau),8\tau,8\tau$&$/+1/\stackrel{\tau = -1/2}{\rightarrow}{\rm max}\stackrel{\tau = 0}{\rightarrow}/-1/$\\
\hline
L$(\pi/2,\pi/2,\pi/2)$ & $-6\tau$ & $ 4(1 + 2\tau), 4(1 + 2\tau),8(\tau - 1)$&${\rm max}\stackrel{\tau = -1/2}{\rightarrow}/+1/\stackrel{\tau = +1}{\rightarrow}{\rm min}$\\
\hline
${\varDelta^\ast}\left(0,0, k^{\rm fcc}_{\varDelta^\ast}\right)$ & $-2\tau^{-1}(2 + 2\tau - \tau^2)$ & $a^{\rm fcc}_{\rm \varDelta^\ast1},a^{\rm fcc}_{\rm \varDelta^\ast1},a^{\rm fcc}_{\rm \varDelta^\ast2}$ & $/-1/, \tau > +1$; $/+1/, \tau < -1$ \\
\hline
${\varSigma^\ast}\left(0,k^{\rm fcc}_{\varSigma^\ast},k^{\rm fcc}_{\varSigma^\ast}\right)$ & $4(1 - 2\tau)^{-1}-2\tau$& $a^{\rm fcc}_{\rm \varSigma^\ast1},a^{\rm fcc}_{\rm \varSigma^\ast2},a^{\rm fcc}_{\rm \varSigma^\ast3}$ & $/+1/, \tau > +1$; $/-1/, \tau < 0$\\
\hline
\end{tabular}
\end{table*}

Consider exact expressions for the density of states in the sc and bcc lattice within the tight-binding approximation (see main text).
\begin{figure*}[!t]
\centering
\includegraphics[angle=-90,width=0.9\textwidth]{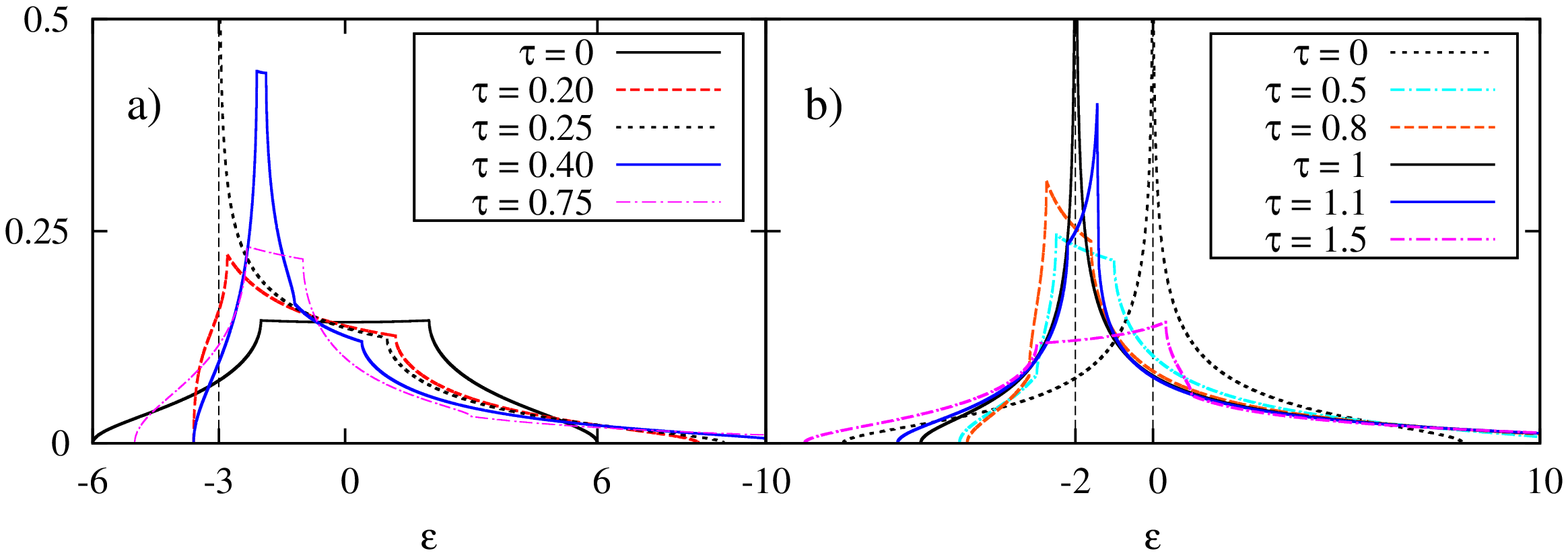}
\caption{DOS $t\rho(\epsilon)$ for sc (a) and bcc (b) lattices at different $\tau$. Vertical dashed lines show the position of giant van Hove singularities.
\label{fig:DOS_vs_e}
}
\end{figure*}
The DOS can be presented as a sum of three contributions
\begin{equation}\label{eq:DOS_exp}
	\rho(\epsilon; \tau) =  \mathcal{R}_\psi(\epsilon, \tau) + \mathcal{R}_{\varphi'}(\epsilon, \tau) + \mathcal{R}_\varphi(\epsilon, \tau).
\end{equation}

\begin{figure*}[hb!]
\centering
\includegraphics[angle=-90,width=0.8\textwidth]{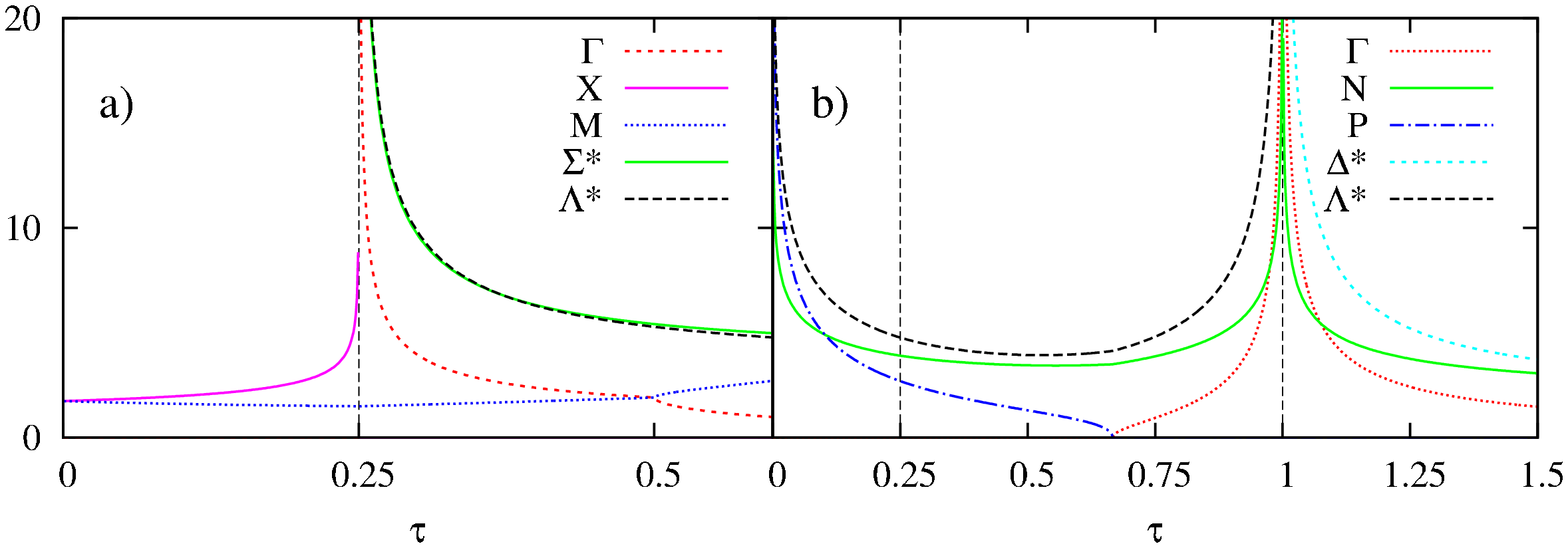}
\caption{$\tau$ dependence of DOS $W\rho(\epsilon;\tau)$
at each van Hove singularity level for sc (a) and bcc (b) lattices ($W$ is the bandwidth). In the case(a), DOS values at $w^{\rm sc}_{\varLambda^\ast}$ and $w^{\rm sc}_{\varSigma^\ast}$ are almost equal, see~main text.
\label{fig:DOS_vs_tau}
}
\end{figure*}

For sc lattice, depending on $\tau$ value the contributions $\mathcal{R}^{\rm sc}_{i}$ read   
\\
1. $\tau \le 1/4$. $\mathcal{R}^{\rm sc}_{\psi} = \mathcal{R}^{\rm sc}_{\varphi'} = 0$  
\begin{equation}\label{eq:R0_tau<=1/4}
	\mathcal{R}^{\rm sc}_\varphi =
	\begin{cases}
		\Phi_{\rm sc}(x^{\rm sc}_{\psi1}, +1),& w^{\rm sc}_\Gamma < \epsilon < w^{\rm sc}_{\rm X}(\tau)\\
		\Phi_{\rm sc}(-1, +1),& w^{\rm sc}_{\rm X}(\tau) < \epsilon < w^{\rm sc}_{\rm M}(\tau)\\
		\Phi_{\rm sc}(-1, x^{\rm sc}_{\psi2}),& w^{\rm sc}_{\rm M}(\tau) < \epsilon < w^{\rm sc}_{\rm R}.
	\end{cases}		
\end{equation}

2. $1/4 < \tau\le1/2$.
\begin{eqnarray}\label{eq:R1_sc}
	\mathcal{R}^{\rm sc}_\psi  &=&
	\begin{cases}
	2\Psi_{\rm sc}(x^{\rm sc}_{\varphi},+1),& w^{\rm sc}_{\rm X}<\epsilon < w^{\rm sc}_{\varSigma^\ast}\\
	2\Psi_{\rm sc}(x^{\rm sc}_{\zeta -},x^{\rm sc}_{\zeta +}),& w^{\rm sc}_{\varSigma^\ast}< \epsilon < w^{\rm sc}_{\varLambda^\ast},
	\end{cases}\\
\label{eq:R2_sc}
	\mathcal{R}^{\rm sc}_{\varphi'}  &=&
	\begin{cases}
	2[\Phi_{\rm sc}(x^{\rm sc}_{\psi1}, x^{\rm sc}_{\zeta -}) +\\+ \Phi_{\rm sc}(x^{\rm sc}_{\zeta +}, +1)], w^{\rm sc}_{\varSigma^\ast}<\epsilon < w^{\rm sc}_{\varLambda^\ast}\\
	2\Phi_{\rm sc}(x^{\rm sc}_{\psi1}, +1), w^{\rm sc}_{\varLambda^\ast}< \epsilon < w^{\rm sc}_\Gamma,
 	\end{cases}\\
\label{eq:R0_sc}
	\mathcal{R}^{\rm sc}_\varphi  &=&
	\begin{cases}	
	\Phi_{\rm sc}(-1,x^{\rm sc}_{\psi1}),& w^{\rm sc}_{\rm X}<\epsilon < w^{\rm sc}_\Gamma \\
	\Phi_{\rm sc}(-1,+1),& w^{\rm sc}_\Gamma<\epsilon < w^{\rm sc}_{\rm M} \\
	\Phi_{\rm sc}(-1,x^{\rm sc}_{\psi2}),& w^{\rm sc}_{\rm M}<\epsilon < w^{\rm sc}_{\rm R}.
	\end{cases}
\end{eqnarray}	
3. $\tau>1/2$. The kinks of the functions $\mathcal{R}^{\rm sc}_{\varphi}$ and $\mathcal{R}^{\rm sc}_{\varphi'}$ at $\epsilon =  w_0^{\rm sc}(\tau) = 4\tau - \tau^{-1}$ cancel each other.
\begin{eqnarray}\label{eq:R1_sc2}
	\mathcal{R}^{\rm sc}_\psi &=&	
	\begin{cases}
	2\Psi_{\rm sc}\left(x^{\rm sc}_{\varphi}, +1\right),& w^{\rm sc}_{\rm X}<\epsilon < w^{\rm sc}_{\varSigma^\ast}\\
	2\Psi_{\rm sc}\left(x^{\rm sc}_{\zeta -},x^{\rm sc}_{\zeta +}\right),& w^{\rm sc}_{\varSigma^\ast}<\epsilon <  w^{\rm sc}_{\varLambda^\ast},
	\end{cases}\\
\label{eq:R2_sc2}
	\mathcal{R}^{\rm sc}_{\varphi'} &=&	
	\begin{cases}
	2\Phi_{\rm sc}\left(x^{\rm sc}_{\zeta +}, {\rm min}[x^{\rm sc}_{\psi2},+1]\right)\\ + 2\Phi_{\rm sc}\left(x^{\rm sc}_{\psi1}, x^{\rm sc}_{\zeta -}\right),  w^{\rm sc}_{\varSigma^\ast}<\epsilon <  w^{\rm sc}_{\varLambda^\ast}\\
	2\Phi_{\rm sc}\left(x^{\rm sc}_{\psi1},{\rm min}[x^{\rm sc}_{\psi2},+1]\right),  w^{\rm sc}_{\varLambda^\ast}<\epsilon <  w^{\rm sc}_0,
	\end{cases}\\
\label{eq:R0_sc2}
	\mathcal{R}^{\rm sc}_\varphi &=&
	\begin{cases}
	\Phi_{\rm sc}\left(-1, x^{\rm sc}_{\psi1}\right),& w^{\rm sc}_{\rm X}<\epsilon < w^{\rm sc}_{\rm M}\\
	\Phi_{\rm sc}\left(-1,x^{\rm sc}_{\psi1}\right) \\ \hspace{0.5cm}+ \Phi_{\rm sc}\left(x^{\rm sc}_{\psi2},+1\right),& w^{\rm sc}_{\rm M}< \epsilon <  w^{\rm sc}_0\\
	\Phi_{\rm sc}\left(-1,x^{\rm sc}_{\psi2}\right) \\ \hspace{0.5cm}+ \Phi_{\rm sc}\left(x^{\rm sc}_{\psi1},+1\right),& w^{\rm sc}_0< \epsilon <  w^{\rm sc}_\Gamma\\
	\Phi_{\rm sc}\left(-1,x^{\rm sc}_{\psi2}\right),& w^{\rm sc}_\Gamma<\epsilon < w^{\rm sc}_{\rm R}.
	\end{cases}
\end{eqnarray}
In these equations $F(x, y) = \mathbb{K}(1 - y/x)/\sqrt{x}$ is symmetric function, which is infinite when $x$ or $y$ tends to zero, 
$$
\mathbb{K}(m) = \int_0^{\pi/2} \frac{d\phi}{\sqrt{1 - m\sin^2\phi}}
$$
being full elliptic integral of the first kind 
\begin{equation}
\label{eq:SC:Psi_def}
	\Psi_{\rm sc}(x_1, x_2; \epsilon, \tau) = \frac2{\pi^3}\int\limits_{x_1}^{x_2}\frac{dx\,F(-\zeta_{\rm sc}(x; \epsilon, \tau), \psi_{\rm sc}(x; \epsilon, \tau) )}{\sqrt{1 - x^2}},
\end{equation}
\begin{equation}
\label{eq:BCC:Psi_def}
	\Phi_{\rm sc}(x_1, x_2; \epsilon, \tau) = \frac2{\pi^3}\int\limits_{x_1}^{x_2}\frac{dx\,F(\zeta_{\rm sc}(x; \epsilon, \tau), \varphi_{\rm sc}(x; \epsilon, \tau) )}{\sqrt{1 - x^2}},
\end{equation}
where
\begin{eqnarray}
	\zeta_{\rm sc}(x; \epsilon, \tau) &=&  16(\tau(\epsilon  + 2x) + (1 - 2\tau x)^2), \\
	\varphi_{\rm sc}(x; \epsilon, \tau) &=& (\epsilon + 2x + 4\tau)^2, \\
	\psi_{\rm sc}(x; \epsilon, \tau) & = & (\epsilon + 2x - 4\tau)^2 - 16(1 - 2\tau x)^2,
\end{eqnarray}
the $x$ integration bounds are 
$x^{\rm sc}_{\psi s}(\epsilon, \tau) = (1/2)({4((-1)^s + \tau) - \epsilon})/({1 + 4(-1)^s\tau})$, $s = 1,2$
$x^{\rm sc}_\varphi(\epsilon, \tau) = -2\tau - \epsilon/2$,
$x^{\rm sc}_{\zeta\pm}(\epsilon,\tau) = (\frac12 \pm \sqrt{-\frac34 - \tau \epsilon})/2\tau$, at $\epsilon < w^{\rm sc}_{\varLambda^\ast}(\tau)$.
Plots for DOS for different values of $\tau$ for sc lattice are presented in Fig.~\ref{fig:DOS_vs_e}a.

For bcc lattice we present the following form for the contributions
$\mathcal{R}^{\rm bcc}_{i}$

\noindent 1. $\tau \le 1$. 
\begin{eqnarray}\label{eq:R1_bcc_small_tau}
	\mathcal{R}^{\rm bcc}_\psi &=& \begin{cases}
	2\Psi_{\rm bcc}(0, x^{\rm bcc}_{\varphi}),& w^{\rm bcc}_{\rm P} < \epsilon < w^{\rm bcc}_{\varLambda^\ast}\\
	2\Psi_{\rm bcc}(0, x^{\rm bcc}_{\zeta}),& w^{\rm bcc}_{\varLambda^\ast} < \epsilon < w^{\rm bcc}_{\rm N},
	\end{cases}\\
\label{eq:R2_bcc_small_tau}
	\mathcal{R}^{\rm bcc}_{\varphi'} &=& \begin{cases}
	2\Phi_{\rm bcc}(x^{\rm bcc}_{\zeta}, x^{\rm bcc}_{\psi1}),& w^{\rm bcc}_{\varLambda^\ast} < \epsilon < w^{\rm bcc}_{\rm N}\\
	2\Phi_{\rm bcc}(0, x^{\rm bcc}_{\psi1}),& w^{\rm bcc}_{\rm N} < \epsilon < w^{\rm bcc}_{0},
	\end{cases}\\
\label{eq:R0_bcc_small_tau}
	\mathcal{R}^{\rm bcc}_\varphi &=& \Phi_{\rm bcc}(x^{\rm bcc}_{\psi1}, +1),
\end{eqnarray}
2. $\tau>1$. 
\begin{eqnarray}
\label{eq:R1_bcc_large_tau}
	\mathcal{R}^{\rm bcc}_\psi &=&
	\begin{cases}	
	2\Psi_{\rm bcc}(0, x^{\rm bcc}_\varphi),& w^{\rm bcc}_{\rm P} < \epsilon < w^{\rm bcc}_{\rm N}\\
	2\Psi_{\rm bcc}(x^{\rm bcc}_\zeta, +1),& w^{\rm bcc}_{\rm N} < \epsilon <  w^{\rm bcc}_{\varDelta^\ast},	
	\end{cases}\\
\label{eq:R2_bcc_large_tau}
	\mathcal{R}^{\rm bcc}_{\varphi'} &=&
	\begin{cases}	
	2\Phi_{\rm bcc}(0, x^{\rm bcc}_{\zeta}),& w^{\rm bcc}_{\rm N} < \epsilon < w^{\rm bcc}_{\varDelta^\ast}\\
	2\Phi_{\rm bcc}(0, x^{\rm bcc}_{\psi1}) \\\hspace{0.5cm}+ 2\Phi_{\rm bcc}(x^{\rm bcc}_{\psi2}, +1),& w^{\rm bcc}_{\varDelta^\ast} < \epsilon < w^{\rm bcc}_\Gamma.	
	\end{cases}\\
\label{eq:R0_bcc_large_tau}
	\mathcal{R}^{\rm bcc}_\varphi &=&
\begin{cases}	
	\Phi_{\rm bcc}(x^{\rm bcc}_{\psi1}, x^{\rm bcc}_{\psi2}),& w^{\rm bcc}_{\varDelta^\ast} < \epsilon < w^{\rm bcc}_\Gamma\\
	\Phi_{\rm bcc}(x^{\rm bcc}_{\psi1}, +1),& w^{\rm bcc}_\Gamma < \epsilon < w^{\rm bcc}_{\rm H},
\end{cases}
\end{eqnarray}
where arguments $\epsilon, \tau$ are omitted for brevity.  
The kinks of the functions $\mathcal{R}^{\rm bcc}_{\varphi}, \mathcal{R}^{\rm bcc}_{\varphi'}$ at $\epsilon = w_0^{\rm bcc} = 2\tau$ cancel each other.
\begin{equation}\label{eq:Psi_bcc_def}
	\Psi_{\rm bcc}(x_1, x_2; \epsilon, \tau) = \frac2{\pi^3}\int\limits_{x_1}^{x_2}\frac{dxF(-\zeta_{\rm bcc}(x; \epsilon, \tau), \psi_{\rm bcc}(x; \epsilon, \tau) )}{\sqrt{x(1 - x)}},
\end{equation}
\begin{equation}\label{eq:Phi_bcc_def}
	\Phi_{\rm bcc}(x_1, x_2; \epsilon, \tau) = \frac2{\pi^3}\int\limits_{x_1}^{x_2}\frac{dxF(\zeta_{\rm bcc}(x; \epsilon, \tau), \varphi_{\rm bcc}(x; \epsilon, \tau) )}{\sqrt{x(1 - x)}},
\end{equation}
where
\begin{eqnarray}
	\zeta_{\rm bcc}(x; \epsilon, \tau) &=& 16(\tau(\epsilon  + 2\tau) + 4(1 - \tau^2)x), \\
	\varphi_{\rm bcc}(x; \epsilon, \tau) &=& (6\tau + \epsilon - 4\tau x)^2,\\
	\psi_{\rm bcc}(x; \epsilon, \tau) &=& (\epsilon - 2\tau -  4\tau x)^2 - 64x,
\end{eqnarray}
and the $x$ integration boundaries read
$x^{\rm bcc}_{\varphi}(\epsilon,\tau) = (\epsilon + 6\tau)/{4\tau}$, at $w^{\rm bcc}_{\rm P} < \epsilon$,
$x^{\rm bcc}_\zeta(\epsilon,\tau) = {\tau(\epsilon + 2\tau)}/{4(\tau^2 - 1)}$, at $(\epsilon - w^{\rm bcc}_\zeta)(\tau - 1) > 0$,
$x^{\rm bcc}_{\psi1,2}(\epsilon,\tau) = {\left(\mp2 + \sqrt{\tau(\epsilon - 2\tau) + 4}\right)^2}/(4\tau^2)$, at $\epsilon > w^{\rm bcc}_{\varDelta^\ast}$.
DOS plots for different values of $\tau$ for bcc lattice are shown in Fig.~\ref{fig:DOS_vs_e}b.

Full understanding for the dependence of DOS on $\tau$ can be obtained using the $\tau$ dependence of $\rho(w^{\rm vHS}_i(\tau), \tau)$ at the levels of van Hove points $\epsilon = w^{\rm vHS}_i(\tau)$.
These plots are shown in Fig.~\ref{fig:DOS_vs_tau}.
For sc lattice, the maximal value of DOS at $\tau > 1/4$ is achieved at the levels $\epsilon = w^{\rm sc}_{\varSigma^\ast}$ and $w^{\rm sc}_{\varLambda^\ast}$.
These values slowly decrease as $\tau$ is shifted from 1/4 ($\rho_{\rm sc}(w^{\rm sc}_{\varSigma^\ast}(\tau), \tau) \approx \rho_{\rm sc}(w^{\rm sc}_{\varSigma^\ast}(\tau), \tau) \sim (\tau - 1/4)^{-1/2}$).
For bcc lattice, the maximal value of DOS is always achieved at the energy level corresponding to inner points of the Brillouin zone:
$\epsilon = w^{\rm bcc}_{\varLambda^\ast}$ at $\tau < 1$, $\epsilon = w^{\rm bcc}_{\varDelta^\ast}$ at $\tau > 1$.

\end{document}